\documentclass[iop,apj]{emulateapj}

\usepackage{apjfonts}
\usepackage{natbib}
\usepackage{amssymb}
\usepackage{enumitem}
\usepackage{multirow}
\usepackage[breaklinks,colorlinks,citecolor=blue,linkcolor=magenta]{hyperref}

\newcommand{\OVIdblt}{{\rm O}\kern 0.1em{\sc vi}~$\lambda\lambda 1031, 1037$} 
\newcommand{\MgIIdblt}{{\rm Mg}\kern 0.1em{\sc ii}~$\lambda\lambda 2796, 2803$}
\newcommand{\OVI}{\hbox{{\rm O}\kern 0.1em{\sc vi}}}
\newcommand{\OVII}{\hbox{{\rm O}\kern 0.1em{\sc vii}}}
\newcommand{\OVIII}{\hbox{{\rm O}\kern 0.1em{\sc viii}}}
\newcommand{\MgII}{\hbox{{\rm Mg}\kern 0.1em{\sc ii}}}
\newcommand{\NeVIII}{\hbox{{\rm Ne}\kern 0.1em{\sc viii}}}
\newcommand{\MgX}{\hbox{{\rm Mg}\kern 0.1em{\sc x}}}
\newcommand{\CIV}{\hbox{{\rm C}\kern 0.1em{\sc iv}}}
\newcommand{\HI}{\hbox{{\rm H}\kern 0.1em{\sc i}}}
\newcommand{\kms}{\hbox{km~s$^{-1}$}}
\newcommand{\etal}{et~al.}

\newcommand{\vfifty}{\hbox{$\Delta v(50)$}}
\newcommand{\vninety}{\hbox{$\Delta v(90)$}}

\shorttitle{OVI CGM Kinematics}
\shortauthors{\sc Pointon {\etal}}
\slugcomment{Accepted to ApJ, June 2, 2017}

\begin{document}

\title{The Impact of the Group Environment on the {\OVI} Circumgalactic Medium}
\author{
Stephanie K. Pointon$^{1}$
}
\author{
Nikole M. Nielsen$^{1}$
}
\author{Glenn G. Kacprzak$^{1}$
}
\author{Sowgat Muzahid$^{2}$
}
\author{Christopher W. Churchill$^{3}$
}
\author{Jane C. Charlton$^{4}$
}

\affil{$^1$ Centre for Astrophysics and Supercomputing, Swinburne
  University of Technology, Hawthorn, Victoria 3122, Australia;
  spointon@swin.edu.au
}
\affil{$^2$ Leiden Observatory, University of Leiden, PO Box 9513, NL-2300 RA Leiden, The Netherlands
}
\affil{
	$^3$ Department of Astronomy, New Mexico State University, Las Cruces, NM 88003, USA
}
\affil{$^4$ Department of Astronomy and Astrophysics, The Pennsylvania State University, State College, PA 16801, USA
}

\begin{abstract}

We present a study comparing {\OVIdblt} doublet absorption found towards group galaxy environments with that of isolated galaxies. The {\OVI} absorption in the circumgalactic medium (CGM) of isolated galaxies has been studied previously by the "Multiphase Galaxy Halos" survey, where the kinematics and absorption properties of the CGM have been investigated. We extend these studies to group environments. We define a galaxy group to have two or more galaxies having a line-of-sight velocity difference of no more than 1000~{\kms} and located within 350 kpc (projected) of a background quasar sightline. We identified a total of six galaxy groups associated with {\OVI} absorption $W_{\rm r}>0.06$~{\AA} that have a median redshift of $\langle z_{\rm gal} \rangle = 0.1669$ and a median impact parameter of $\langle D \rangle = 134.1$~kpc. An additional 12 non-absorbing groups were identified with a median redshift of $\langle z_{\rm gal} \rangle = 0.2690$ and a median impact parameter of $\langle D \rangle = 274.0$~kpc. We find the average equivalent width to be smaller for group galaxies than for isolated galaxies $(3\sigma)$. However, the covering fractions are consistent with both samples. We used the pixel-velocity two-point correlation function method and find that the velocity spread of {\OVI} in the CGM of group galaxies is significantly narrower than that of isolated galaxies $(10\sigma)$. We suggest that the warm/hot CGM does not exist as a superposition of halos, instead, the virial temperature of the halo is hot enough for {\OVI} to be further ionised. The remaining {\OVI} likely exists at the interface between hot, diffuse gas and cooler regions of the CGM.

\end{abstract}


\section{Introduction}
\label{sec:intro}

The environment in which galaxies exist plays a significant part in the way they will evolve. While unperturbed isolated galaxies have a disk-like structure, interactions between galaxies can result in other interesting phenomena such as tidal streams, shells, and increased star formation rates \citep[e.g.,][]{barnes92,veilleux05, poggianti16}. In the most extreme cases interactions remove the galaxy gas reservoir, leading to the quenching of star formation, and the original disk-like structures of the merging galaxies deform into elliptical galaxies \citep[e.g.,][]{gunn72,cowie77, larson80, nulsen82,moore96,cen99, oppenheimer08, lilly-bathtub}. These phenomena are observed in images taken of interacting galaxies with, for example, bursts of star formation activity \citep[e.g.,][]{keel85,barnes04} or streams of HI gas connecting galaxies can be seen \citep[e.g.,][]{bridge10}. However, we know very little about the impact of a merger or interaction on the diffuse gas around galaxies.

Of particular interest is the circumgalactic medium (CGM), which is a vast halo of diffuse gas surrounding galaxies out to radii $\approx200$~kpc \citep[e.g.,][]{kacprzak08,chen10a,steidel10,kacprzak11morph,tumlinson11,rudie12,burchett13,magiicat1,magiicat2,werk13}. The mass of this halo is comparable to the gas mass of the galaxy itself \citep[i.e., the ISM;][]{thom11,tumlinson11,werk13} and hence plays an important role in the evolution of galaxies. Models of the CGM indicate that gas can flow in and out of this reservoir, which in turn controls the star formation rate of the galaxy and the metallicity of stars formed from this gas \citep{oppenheimer08, lilly-bathtub,kacprzak16}. While a more concrete model of how this gas drives isolated galaxy evolution is being built, we are still only just beginning to study the effects an interaction or merger can have on the CGM.
  
Since the visible components of galaxies are clearly affected by galaxy interactions, it is reasonable to expect that such effects would also take place in the CGM. Indeed, due to the large radius of the CGM, two galaxies could have overlapping gaseous halos even if they do not yet show any signs that an interaction is taking place \citep{tully09,bordoloi11,stocke14}. Thus, it is possible to use the CGM to understand the very first processes that take place in an interaction. Similarly, as interactions often result in the ISM being stripped from the galaxy leading to gas depletion \citep[e.g.,][]{gunn72,larson80,fujita99}, the CGM could also be susceptible to gas removal through tidal interactions \citep{chen09}. This, in turn, would affect the ability of galaxies to form stars later in life as accretion of gas from the CGM is what sustains star formation. 

Previous studies of the CGM in group environments have used the {\MgIIdblt} absorption doublet. \citet{bordoloi11} stacked background galaxy spectra and found that {\MgII} absorption at a given equivalent width was located at higher impact parameters in group environments compared to isolated environments. This could be explained by constructing a simple model where the CGM of the group galaxies were superimposed. Thus the extended CGM is due to more gas along the line of sight, and hence, the {\MgII} component of the CGM is minimally affected by the interactions in the group. Recent work by Nielsen {\etal} (2017, in prep) has also compared {\MgII} absorption between group and isolated environments. They found that the absorption associated with group galaxies had larger equivalent widths than for isolated environments. The authors also found an increased fraction of {\MgII} components with higher velocities in group environments, which indicated that there could be some interaction between the CGM halos of the group galaxies.

Another ideal absorption doublet to trace the CGM is {\CIV}, which traces gas with temperatures $T\approx10^{4-5}$~K. \citet{burchett16} compared {\CIV} absorption in isolated and group/cluster environments and found that for a massive and dense  group environment with seven group member galaxies ($M_h > 10^{12.7} M_{\odot}$ and ${M}_r \leq -19$), no {\CIV} was detected despite associated {\HI} absorbers. They find that the cause of {\CIV} depletion in group environments is not clear due to the limited sample size. However, if larger studies of {\CIV} absorption follow similar trends, the authors suggest that the CGM of the group galaxies may be experiencing ram pressure or tidal stripping, and that {\CIV} traces gas with higher temperatures than {HI} absorbers. 

The warmer, more diffuse phase of the CGM gas is traced by {\OVI} due to its higher ionisation state and temperature $T\approx10^{5.5}$~K. {\OVI} can be both collisionally ionised and photo-ionised, which makes interpretation more difficult, because it traces multiphase structures \citep{mo96,maller04,dekel06}. Previous studies by \citet{stocke14,stocke17} compared {\OVI} absorption in group environments to that found in isolated environments. They found that {\OVI} absorption profiles had to be modeled using fewer, broader components in group environments, indicating a warmer environment. However, the total absorption profile was narrower from which the authors reasoned that {\OVI} could not be distributed over the "circum-group" medium, especially since {\OVI} is unstable due to rapid cooling. This suggests that a diffuse halo of warm CGM gas would be difficult to maintain, and that {\OVI} exists at the interface of the diffuse, hot ($T>10^6$~K) and cooler, photoionized regions embedded in the "circum-group" medium. The existence of interfaces between hot and cold gas within the CGM has also been found by \citet{churchill12,churchill13} and \citet{stern16} for isolated environments.

\begin{deluxetable*}{llllclc}
	\tablecolumns{7}
	\tablewidth{0pt}
	\setlength{\tabcolsep}{0.06in}
	\tablecaption{Quasar Observations \label{tab:obsqso}}
	\tablehead{
		\colhead{J-Name}           	&
		\colhead{RA (J2000)}           		&
		\colhead{DEC (J2000)}           	&
		\colhead{$z_{\rm qso}$}     	&
		\colhead{Instrument}			&
		\colhead{REF\tablenotemark{a}}	&			
		\colhead{COS/FUSE PID}}
	\startdata
	J004706$+$031955	&	$00:47:05.93$	&	$+03:19:54.90$	&	$0.6233$	&	COS	                        &$(3)$  &$12275$\\
J012529$-$000556	&	$01:25:28.84$	&	$-00:05:55.93$	&	$1.0748$	&	COS                             &$(1)$  &$13398$\\
J022815$-$405714	&	$02:28:15.17$	&	$-40:57:14.29$	&	$0.4934$	&	COS	                        &$(2)$  &$11541$\\
J035129$-$142909 	&	$03:51:28.54$	&	$-14:29:08.71$	&	$0.6163$	&	COS	                        &$(5)$  &$13398$\\
J040748$-$121137	&	$04:07:48.43$	&	$-12:11:36.66$	&	$0.5726$	&	COS, FUSE\tablenotemark{b}	&$(4)$  &$11541$, B$087$\\
J045609$-$215909	&	$04:56:08.92$	&	$-21:59:09.40$	&	$0.5335$	&	COS	                        &$(7)$  &$13398$\\
J092838$+$602521	&	$09:28:37.98$	&	$+60:25:21.02$	&	$0.2959$	&	COS	                        &$(3)$  &$11598$\\
J111909$+$211918	&	$11:19:08.68$	&	$+21:19:18.01$	&	$0.1765$	&	FUSE\tablenotemark{b}           &$(3)$  &P$101 $\\
J113328$+$032719	&	$11:33:27.78$	&	$+03:27:19.17$	&	$0.5245$	&	COS	                        &$(6)$  &$11598$\\
J113910$-$135044	&	$11:39:10.70$	&	$-13:50:43.64$	&	$0.5565$	&	COS	                        &$(1)$  &$12275$\\
J130112$+$590206        &       $13:01:12.93$   &       $+59:02:06.75$  &       $0.4778$        &	COS	                        &$(8)$	&$11541$\\
J131956$+$272808        &       $13:19:56.23$   &       $+27:28:08.22$  &       $1.0147$        &	COS	                        &$(8)$	&$11667$\\
J135704$+$191907        &       $13:57:04.43$   &       $+19:19:07.37$  &       $0.7200$        &	COS	                        &$(8)$	&$13398$\\
J170441$+$604430        &       $17:04:41.37$   &       $+60:44:30.50$  &       $0.3719$        &	COS	                        &$(8)$	&$12276$\\
J182157$+$642037        &       $18:21:57.31$   &       $+64:20:36.37$  &       $0.2970$        &	COS	                        &$(8)$	&$12038$\\[-5pt]

	\enddata
	\tablenotetext{a}{Reference for the position and redshift of the quasars are: $(1)$ \citet{beasley02}, $(2)$ \citet{beuermann99}, $(3)$ \citet{zickgraf03}, $(4)$ \citet{fey04}, $(5)$ \citet{li96}, $(6)$ \citet{wu12}, $(7)$ \citet{healey07}, $(8)$ \cite{chen01b}}
	\tablenotetext{b}{These quasar spectra were obtained from Wakker (2016, private communication).}
\end{deluxetable*}

Recently, \citet{oppenheimer16} used the EAGLE simulations to compare the column density of various oxygen ionisation states as a function of halo mass and found that the presence and strength of {\OVI} absorption are strongly related to the mass of the halo. Halos of sub-$L^*$ galaxies contain little {\OVI} absorption because the virial temperature is too low to ionise oxygen to such a high state. As the halo mass increases, the presence and strength of {\OVI} absorption increases until the virial temperature approaches the temperature where the ionisation fraction of {\OVI} peaks, which occurs in the CGM surrounding $L^*$ galaxies. Finally, as the halo mass increases and virial temperature beyond this point, the presence of {\OVI} absorption decreases as oxygen is ionised to higher states. They find that oxygen is more likely to exist as {\OVII} and {\OVIII} for more massive halos, in particular, halos significantly larger than $L^*$. 

To investigate the simulation predictions from \citet{oppenheimer16} and compare to {\MgII} and CIV absorption results, we use two {\OVI} samples, one which contains only isolated galaxies and the other contains galaxies in group environments. The isolated sample, which represents $L^*$ galaxies, has been presented in the "Multiphase Galaxy Halos" survey \citep{kacprzak15,muzahid15,muzahid16,nielsenovi}. Here we present the group sample, representing halos larger than that of $L^*$ environments, which is the combination of data from the "Multiphase Galaxy Halos" survey and literature \citep{ellingson91,chen01b,chen09,prochaska11,werk12,werk13,werk14,johnson13,savage14}. A combination of \textit{HST} images and \textit{HST}/COS spectra allows us to investigate the equivalent widths, covering fractions, and kinematics of group environments and compare them to existing surveys of {\OVI} absorption towards isolated galaxies.

\begin{figure*}[ht]
	\centering
	\includegraphics[width=\linewidth]{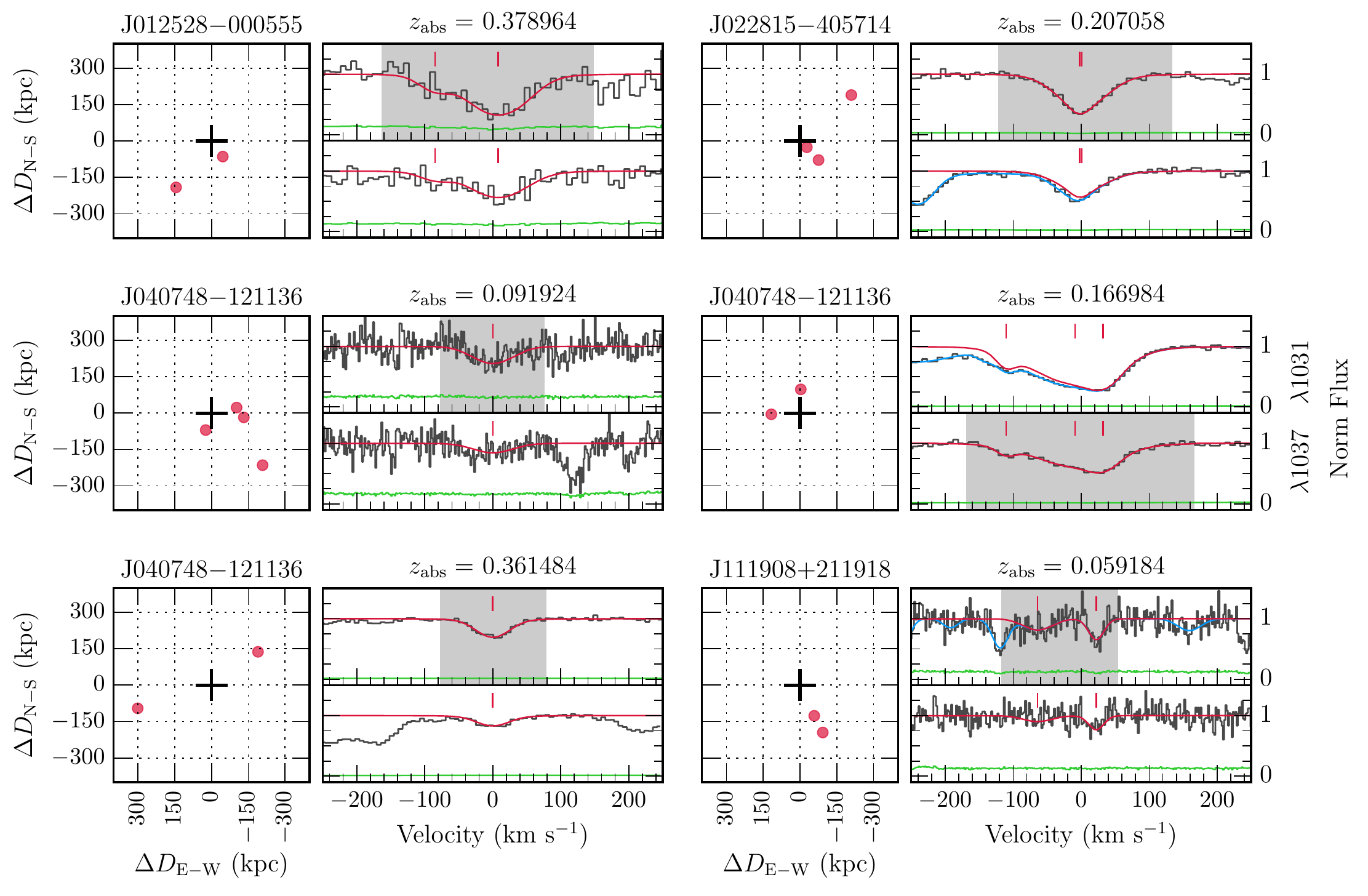}
	\caption[]{Sky locations and {\it HST}/COS or FUSE spectra absorption profiles for each group environment with {\OVI} absorption. The absorbers J040748$-$121136 at $z_{\text{abs}}=0.091917$ and J111908$+$211918 at $z_{\text{abs}}=0.059188$ are from FUSE spectra. The relative positions of the galaxies in each group (red circles) are shown with respect to the locations of the quasar (black cross). The top profile shows the {\OVI}~{$\lambda 1031$} absorption line while the bottom profile shows the {\OVI}~{$\lambda 1037$} line, where the data is shown in black, the error spectrum in green, and the {\OVI} model in red. The center of each model component is indicated by a red tick. The blue line shows the modeled fit to the data where other transitions were modeled in addition to {\OVI} due to the presence of blends. The shaded region contains the absorbing pixels within the velocity regions defined in Section \ref{sec:qso_spec}.}
	\label{fig:fitplot}
\end{figure*}

This paper is organized as follows: In Section \ref{sec:methods} we describe our sample of galaxies, quasars, and the selection criterion used to classify group and isolated environments. We also describe how we obtain a physical description of the CGM in such environments using quasar spectra. We present the results of our analysis with absorption properties and kinematics in Section \ref{sec:results}. In Section \ref{sec:discussion} we discuss the differences between {\OVI} absorption for group and isolated environments and in Section \ref{sec:conclusions} we summarize the work of this paper. We use standard a $\Lambda$CDM cosmology $(H_o = 70\kms Mpc^{-1}, \Omega_M = 0.3, \Omega_\Lambda = 0.7)$.

\section{Sample and Data}
\label{sec:methods}
\label{sec:qso_spec}
We identified group environments in a total of ten fields. All but three were identified in quasar spectra from the Cosmic Origins Spectrograph (COS) on \textit{HST}. The remaining three were from the FUSE telescope, and the spectra were provided by B. Wakker (2016, private communication). The quasar locations, redshifts and the observation PIDs are shown in Table~\ref{tab:obsqso}. The quasar spectra that sample foreground galaxy groups with {\OVI} absorption are shown in columns two and four of Figure~\ref{fig:fitplot} as a black line.
\subsection{Galaxy Properties}
\label{sec:gal_prop}
\begin{figure*}[ht]
	\centering
	\includegraphics[scale=0.80]{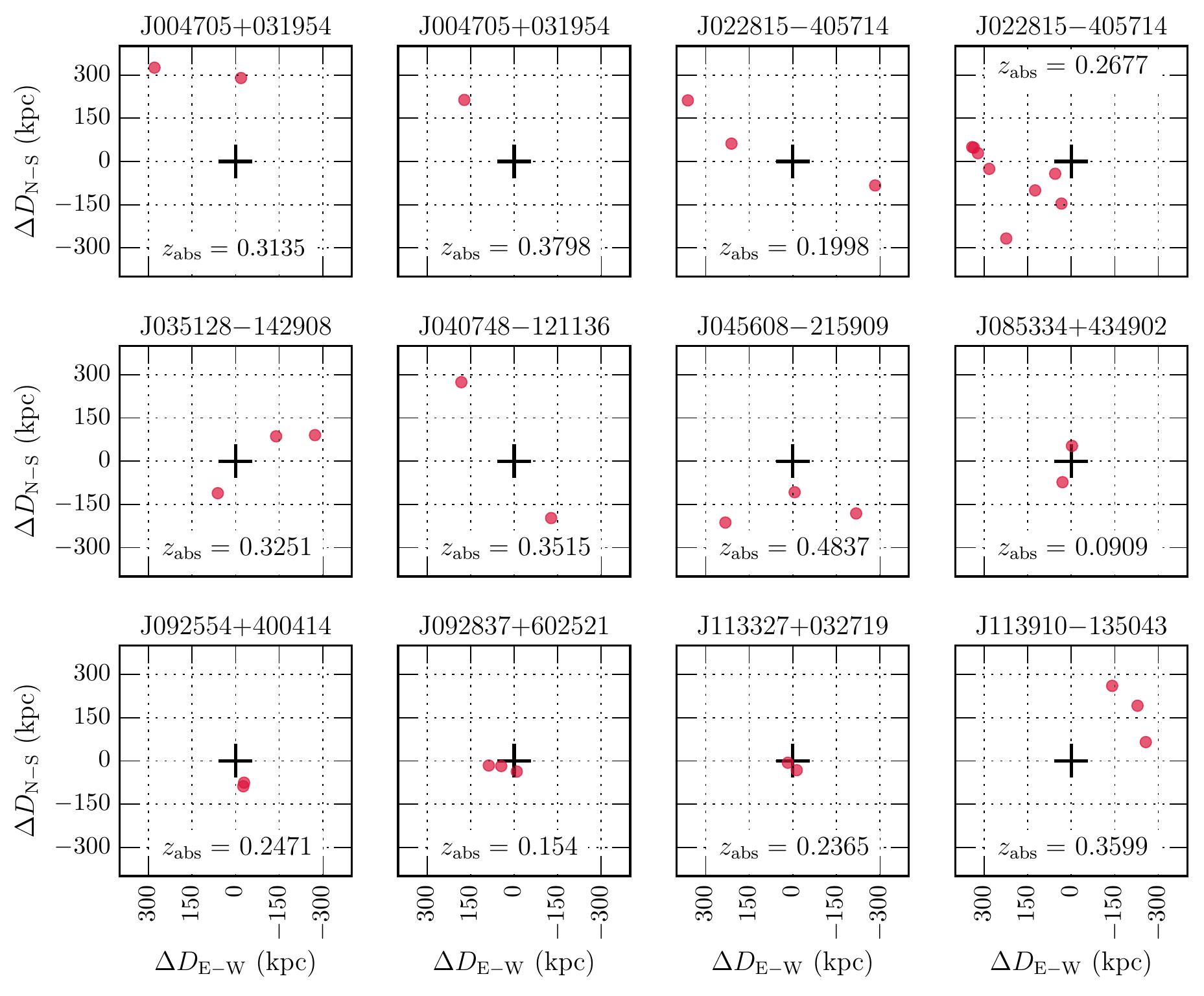}
	\caption[]{Sky locations for each group environment with no associated {\OVI} absorption $W_r < 0.06$~{\AA}. The relative positions of the galaxies in each group (red circles) are shown with respect to the locations of the quasar (black cross).}
	\label{fig:nafitplot}
\end{figure*}

We use a sample of galaxies located in group environments drawn from the "Multiphase Galaxy Halos" survey \citep{kacprzak15,muzahid15,muzahid16,nielsenovi} or obtained from the literature. A group is defined as having the nearest of two or more galaxies within $20$\footnote{The lower limit on the impact parameter is set to match the impact parameter distribution of the isolated sample from \citet{kacprzak15}. to $350$ kpc of the quasar sightline, and the galaxies must have a line-of-sight velocity separation less than $1000$~{\kms}.} With these criteria, we compiled a sample of six group environments with detected {\OVI} absorption above a rest-frame equivalent width of $W_r \geq 0.06 $~{\AA} in background quasar spectra, and an additional 12 with no detected absorption. Two of these group environments detected with {\OVI} absorption were identified by \citet{savage14} from FUSE data. The median redshift and impact parameter of the galaxies in the group environment sample with {\OVI} absorption are $\langle z_{\rm gal} \rangle = 0.1669$ and $\langle D \rangle = 134.1$~{kpc}, respectively. In comparison, the median values for the equivalent width and impact parameter for group environments with no {\OVI} absorption are $\langle z_{\rm gal} \rangle = 0.2690$ and $\langle D \rangle = 274.0$~{kpc}. We show the group galaxy redshift, position and impact parameter in Table \ref{tab:group} as well as the same values for the non-absorbing groups. The values for the right ascension and declination offsets and the angular separation of the galaxies from the quasar sightline, as well at the impact parameter were recalculated using methods from \cite{magiicat1}. In Figure \ref{fig:fitplot} we show the locations of the group galaxy members associated with {\OVI} absorption (red points) with relation to the quasar sightline (central black cross) in the first and third column. Similarly, we show in Figure \ref{fig:nafitplot} the locations of galaxies which were not associated with {\OVI} absorption (red points) relative to the location of the quasar sightline (central black cross).
\begin{deluxetable*}{llrrrrrlccrr}
        \tabletypesize{\footnotesize}
	\tablecolumns{12}
	\tablewidth{0pt}
	\setlength{\tabcolsep}{0.04in}
	\tablecaption{Group Galaxies\label{tab:group}}
	\tablehead{
		\colhead{$(1)$}     	&
		\colhead{$(2)$}     	&
		\colhead{$(3)$}     	&
		\colhead{$(4)$}			&
		\colhead{$(5)$}			&
		\colhead{$(6)$}			&
		\colhead{$(7)$}			&
		\colhead{$(8)$}			&
		\colhead{$(9)$}			&
		\colhead{$(10)$}     	&
		\colhead{$(11)$}     	&
		\colhead{$(12)$}     	\\
		\colhead{Quasar J-Name} &
		\colhead{$z_{\rm gal}$} &
		\colhead{REF\tablenotemark{a}}			&
		\colhead{$\Delta \alpha$}	&
		\colhead{$\Delta \delta$}	&
		\colhead{$\theta$}			&
		\colhead{$D$\tablenotemark{b}}				&
		\colhead{$z_{\rm abs}$}		&
		\colhead{$W_{\rm r}$\tablenotemark{c}}		&
		\colhead{$\log N ({\OVI})$\tablenotemark{c}}	&
		\colhead{$v_{\rm -}$\tablenotemark{c}}	&
		\colhead{$v_{\rm +}$\tablenotemark{c}}\\
		\colhead{}     	&
		\colhead{}     	&
		\colhead{}		&	
		\colhead{$(\arcsec)$}	&
		\colhead{$(\arcsec)$}	&
		\colhead{$(\arcsec)$}	&
		\colhead{(kpc)}			&
		\colhead{}				&
		\colhead{$($\AA$)$}		&
		\colhead{}				&
		\colhead{$($\kms$)$}		&	
		\colhead{$($\kms$)$}		}
	\startdata
	\cutinhead{Properties of Group Galaxies Associated with {\OVI} Absorption $(W_{r}>0.06$~\AA$)$}
                                                      &          &  &          &           &          &         &                               &                                                       &                                                       &                                           &                                                      \\[-12pt]
\rule{0pt}{3ex}\multirow{2}{*}{ J012328$-$000555 }    &$ 0.3787 $&5 &$ -8.8   $&$ -12.3   $&$ 15.07  $&$ 78.3 $ & \multirow{2}{*}{$0.378964$}	& \multirow{2}{*}{$0.32\pm0.04$}			&\multirow{2}{*}{$14.60\pm0.07$}			& \multirow{2}{*}{$-158$\phn}               &\multirow{2}{*}{$143$\phn}                    	   \\
   		 		  		      &$ 0.3792 $&3 &$ 27.7   $&$ -36.5   $&$ 45.80  $&$ 238.1$ 																		 			    									   \\[1.5pt]\hline
                                                      &          &  &          &           &          &         &                               &                                                       &                                                       &                                           &                                                      \\[-12pt]
\rule{0pt}{3ex}\multirow{3}{*}{ J022815$-$405714 }    &$ 0.2065 $&4 &$ -9.1   $&$ -8.4    $&$ 10.87  $&$ 36.8 $ & \multirow{3}{*}{$0.207058$}	& \multirow{3}{*}{$0.199\pm0.007$}			&\multirow{3}{*}{$14.38\pm0.01$}			& \multirow{3}{*}{$-117$\phn}		    &\multirow{3}{*}{$129$\phn}			   	   \\
   				       $ $	      &$ 0.2078 $&4 &$ -24.9  $&$ -25.9   $&$ 32.04  $&$ 108.9$ 																														   \\
   				       $ $            &$ 0.2077 $&4 &$ -69.9  $&$ 63.2    $&$ 82.32  $&$ 279.7$ 																		 			    							 		   \\[1.5pt]\hline
                                                      &          &  &          &           &          &         &                               &                                                       &                                                       &                                           &                                                      \\[-12pt]
\rule{0pt}{3ex}\multirow{5}{*}{ J040748$-$121136 }    &$ 0.0923 $&2 &$ 13.6   $&$ -39.9   $&$ 42.08  $&$ 72.3 $ & \multirow{5}{*}{$0.091924$}	& \multirow{5}{*}{$0.07\pm0.02$}			&\multirow{5}{*}{$13.82\pm0.06$}			& \multirow{5}{*}{$-75$\phn}		    &\multirow{5}{*}{$74$\phn}   		   	   \\
   				       $ $	      &$ 0.0908 $&2 &$ -61.9  $&$ 13.6    $&$ 62.01  $&$ 104.9$                                                                                                                                                                                                         					   \\
   				       $ $	      &$ 0.0914 $&2 &$ -78.9  $&$ -10.6   $&$ 77.89  $&$ 132.6$                                                                                                                                                                                                         					   \\
   				       $ $	      &$ 0.0917 $&2 &$ -123.5 $&$ -127.3  $&$ 175.44 $&$ 299.6$                                                                                                                                                                                                         					   \\
   				       $ $	      &$ 0.0908 $&2 &$ 62.5   $&$ -252.3  $&$ 259.64 $&$ 439.4$                                                                                                                                                                                                         					   \\[1.5pt]\hline
                                                      &          &  &          &           &          &         &                               &                                                       &                                                       &                                           &                                                      \\[-12pt]
\rule{0pt}{3ex}\multirow{2}{*}{ J040748$-$121136 }    &$ 0.1670 $&3 &$ -1.1   $&$ 34.8    $&$ 34.81  $&$ 99.4 $ & \multirow{2}{*}{$0.166984$}	& \multirow{2}{*}{$0.245\pm0.004$\tablenotemark{d}}	&\multirow{2}{*}{$14.70\pm0.01$\tablenotemark{d}}	&\multirow{2}{*}{$-170$\tablenotemark{d}}   &\multirow{2}{*}{$156$\tablenotemark{d}}	   	   \\
   				       $ $	      &$ 0.1669 $&3 &$ 41.3   $&$ -1.8    $&$ 40.36  $&$ 115.2$                                                                                                                                                                                                         					   \\[1.5pt]\hline
                                                      &          &  &          &           &          &         &                               &                                                       &                                                       &                                           &                                                      \\[-12pt]
\rule{0pt}{3ex}\multirow{2}{*}{ J040748$-$121136 }    &$ 0.3614 $&2 &$ -38.5  $&$ 27.9    $&$ 46.87  $&$ 236.4$ & \multirow{2}{*}{$0.361484$}	& \multirow{2}{*}{$0.070\pm0.003$}			&\multirow{2}{*}{$13.82\pm0.01$	}			&\multirow{2}{*}{$-73$\phn}		    &\multirow{2}{*}{$74$\phn}			   	   \\
   				       $ $	      &$ 0.3608 $&2 &$ 61.0   $&$ -19.3   $&$ 62.71  $&$ 315.9$                                                                                                                                                                                                         					   \\[1.5pt]\hline
                                                      &          &  &          &           &          &         &                               &                                                       &                                                       &                                           &                                                      \\[-12pt]
\rule{0pt}{3ex}\multirow{2}{*}{ J111908$+$211918 }    &$ 0.0600 $&1 &$ -48.1  $&$ -104.9  $&$ 117.01 $&$ 135.6$ & \multirow{2}{*}{$0.059184$}	& \multirow{2}{*}{$0.06\pm0.02$}			&\multirow{2}{*}{$13.91\pm0.13$}			&\multirow{2}{*}{$-115$\phn}		    &\multirow{2}{*}{$52$\phn}			           \\
   				       $ $            &$ 0.0594 $&1 &$ -83.1  $&$ -174.6  $&$ 190.99 $&$ 219.3$ 																					    									   \\[-6pt]

	\cutinhead{Properties of Group Galaxies Associated with {\OVI} Non-absorption $(W_r < 0.06$~{\AA}$)$\tablenotemark{e}}
	                                                      &          &  &          &           &          &         &                               &                                                       &                                                       &                                           &                                                      \\[-12pt]
\rule{0pt}{3ex}\multirow{3}{*}{J004705$+$031954 }	&$ 0.3142 $&$ 	6  $&$ 	-4.0    $&$ 63.0       $&$ 63.13       $&$ 290	 $ & \multirow{3}{*}{$0.3135$} & \multirow{3}{*}{$<0.046$} & \multirow{3}{*}{$<13.61$} &\multirow{3}{*}{$\cdots$}&\multirow{3}{*}{$\cdots$}  \\
			  $ $	      	   		&$ 0.3130 $&$ 	6  $&$ 	56.0    $&$ 39.0       $&$ 68.13       $&$ 312   $   		      	      			      		    	      	                                       		     	     \\
			  $ $	      	   		&$ 0.3133 $&$ 	6  $&$ 	61.0    $&$ 71.0       $&$ 93.54       $&$ 429	 $   		      	      			      		    	      	                                       		     	     \\[1.5pt]\hline
                                                      &          &  &          &           &          &         &                               &                                                       &                                                       &                                           &                                                      \\[-12pt]
\rule{0pt}{3ex}\multirow{2}{*}{J004705$+$031954}	&$ 0.3810 $&$ 	6  $&$ 	33.0    $&$ 41.0       $&$ 52.60       $&$ 274	 $ & \multirow{2}{*}{$0.3798$} & \multirow{2}{*}{$<0.054$} & \multirow{2}{*}{$<13.68$} &\multirow{2}{*}{$\cdots$}&\multirow{2}{*}{$\cdots$}  \\
			  $ $				&$ 0.3786 $&$ 	6  $&$ 	-96.0   $&$ 8.0        $&$ 96.17       $&$ 499	 $   		      	      			      		    	      			 		      		     	     \\[1.5pt]\hline
                                                      &          &  &          &           &          &         &                               &                                                       &                                                       &                                           &                                                      \\[-12pt]
\rule{0pt}{3ex}\multirow{3}{*}{J022815$-$405714}	&$ 0.1992 $&$ 	4  $&$ 	84.0    $&$ 24.6       $&$ 68.04       $&$ 220	 $ & \multirow{3}{*}{$0.1998$} & \multirow{3}{*}{$<0.011$} & \multirow{3}{*}{$<12.94$} &\multirow{3}{*}{$\cdots$}&\multirow{3}{*}{$\cdots$}  \\
			  $ $	      	     		&$ 0.1997 $&$ 	4  $&$ 	-110.6  $&$ -32.4      $&$ 89.60       $&$ 296	 $   		      	      			      		    	      	                                       		     	     \\
			  $ $	      	     		&$ 0.2006 $&$ 	4  $&$ 	133.7   $&$ 78.4       $&$ 127.88      $&$ 419	 $   		      	      			      		    	      	                                       		     	     \\[1.5pt]\hline
                                                      &          &  &          &           &          &         &                               &                                                       &                                                       &                                           &                                                      \\[-12pt]
\rule{0pt}{3ex}\multirow{8}{*}{J022815$-$405714}	&$ 0.2678 $&$ 	4  $&$ 	16.9    $&$ -13.0      $&$ 18.21       $&$ 70	 $ & \multirow{8}{*}{$0.2677$} & \multirow{8}{*}{$<0.016$} & \multirow{8}{*}{$<13.11$} &\multirow{8}{*}{$\cdots$}&\multirow{8}{*}{$\cdots$}  \\
			  $ $	      	     		&$ 0.2690 $&$ 	4  $&$ 	8.5     $&$ -36.7      $&$ 37.25       $&$ 150	 $   		      	      			      		    	      	                                       		     	     \\
			  $ $	      	     		&$ 0.2680 $&$ 	4  $&$ 	36.2    $&$ -29.2      $&$ 39.98       $&$ 160	 $   		      	      			      		    	      	                                       		     	     \\
			  $ $				&$ 0.2654 $&$ 	4  $&$ 	93.4    $&$ -8.5       $&$ 71.03       $&$ 284	 $ 		      	      			      		    	      	                                       		     	     \\
			  $ $	      	     		&$ 0.2664 $&$ 	4  $&$ 	105.7   $&$ 9.4        $&$ 80.36       $&$ 323	 $   		      	      			      		    	      	                                       		     	     \\
			  $ $	      	     		&$ 0.2706 $&$ 	4  $&$ 	108.4   $&$ 15.4       $&$ 83.28       $&$ 339	 $   		      	      			      		    	      	                                       		     	     \\
			  $ $	      	     		&$ 0.2662 $&$ 	4  $&$ 	116.0   $&$ 16.9       $&$ 89.24       $&$ 345	 $   		      	      			      		    	      	                                       		     	     \\
			  $ $	      	     		&$ 0.2683 $&$ 	4  $&$ 	61.1    $&$ -72.9      $&$ 86.25       $&$ 349	 $   		      	      			      		    	      	                                       		     	     \\[1.5pt]\hline
                                                      &          &  &          &           &          &         &                               &                                                       &                                                       &                                           &                                                      \\[-12pt]
\rule{0pt}{3ex}\multirow{3}{*}{J035128$-$142908}	&$ 0.3236 $&$ 	3  $&$ 	13.0    $&$ -23.5      $&$ 26.72       $&$ 127	 $ & \multirow{3}{*}{$0.3251$} & \multirow{3}{*}{$<0.032$} & \multirow{3}{*}{$<13.41$} &\multirow{3}{*}{$\cdots$}&\multirow{3}{*}{$\cdots$}  \\
			  $ $	      	      		&$ 0.3244 $&$ 	3  $&$ 	-29.9   $&$ 18.5       $&$ 34.33       $&$ 164	 $   		      	      			      		    	      	                                       		     	     \\
			  $ $	      	      		&$ 0.3273 $&$ 	3  $&$ 	-59.0   $&$ 19.5       $&$ 60.31       $&$ 288	 $   		      	      			      		    	      	                                       		     	     \\[1.5pt]\hline
                                                      &          &  &          &           &          &         &                               &                                                       &                                                       &                                           &                                                      \\[-12pt]
\rule{0pt}{3ex}\multirow{3}{*}{J040748$-$121136}	&$ 0.3506 $&$ 	2  $&$ 	-26.0   $&$ -40.3      $&$ 47.66       $&$ 235	 $ & \multirow{3}{*}{$0.3515$} & \multirow{3}{*}{$<0.004$} & \multirow{3}{*}{$<12.51$} &\multirow{3}{*}{$\cdots$}&\multirow{3}{*}{$\cdots$}  \\
			  $ $	      	     		&$ 0.3521 $&$ 	2  $&$ 	37.0    $&$ 55.7       $&$ 66.40       $&$ 329	 $   		      	      			      		    	      	                                       		     	     \\
			  $ $	      			&$ 0.3517 $&$ 	2  $&$ 	80.5    $&$ -141.3     $&$ 161.78      $&$ 802	 $   		      	      			      		    	      	                                       		     	     \\[1.5pt]\hline	  
                                                      &          &  &          &           &          &         &                               &                                                       &                                                       &                                           &                                                      \\[-12pt]
\rule{0pt}{3ex}\multirow{3}{*}{J045608$-$215909}	&$ 0.4838 $&$ 	3  $&$ 	-1.1    $&$ -18.0      $&$ 18.03       $&$ 108	 $ & \multirow{3}{*}{$0.4837$} & \multirow{3}{*}{$<0.032$} & \multirow{3}{*}{$<13.73$} &\multirow{3}{*}{$\cdots$}&\multirow{3}{*}{$\cdots$}  \\
			  $ $	      			&$ 0.4836 $&$ 	3  $&$ 	-20.0   $&$ -16.6      $&$ 24.88       $&$ 284	 $                          		                 	                                     	 		      		     \\
			  $ $	      			&$ 0.4837 $&$ 	3  $&$ 	40.1    $&$ -36.8      $&$ 52.29       $&$ 315	 $  		            		    	      	       	                             	 		      		     	     \\[1.5pt]\hline		  
                                                      &          &  &          &           &          &         &                               &                                                       &                                                       &                                           &                                                      \\[-12pt]
\rule{0pt}{3ex}\multirow{2}{*}{J085334$+$434902}	&$ 0.0915 $&$ 	4  $&$ 	-1.8    $&$ 40.8       $&$ 40.81       $&$ 53	 $ & \multirow{2}{*}{$0.0909$} & \multirow{2}{*}{$<0.054$} & \multirow{2}{*}{$<13.68$} &\multirow{2}{*}{$\cdots$}&\multirow{2}{*}{$\cdots$}  \\
			  $ $	      			&$ 0.0903 $&$ 	7  $&$ 	14.1    $&$ -34.0      $&$ 35.45       $&$ 79	 $   		      	      			      		    	      	                              	      		     	     \\[1.5pt]\hline
                                                      &          &  &          &           &          &         &                               &                                                       &                                                       &                                           &                                                      \\[-12pt]
\rule{0pt}{3ex}\multirow{2}{*}{J092554$+$400414}	&$ 0.2475 $&$ 	8  $&$ 	-8.0    $&$ -20.8      $&$ 21.64       $&$ 81	 $ & \multirow{2}{*}{$0.2471$} & \multirow{2}{*}{$<0.085$} & \multirow{2}{*}{$<13.91$} &\multirow{2}{*}{$\cdots$}&\multirow{2}{*}{$\cdots$}  \\
			  $ $	      			&$ 0.2467 $&$ 	8  $&$ 	-7.2    $&$ -24.1      $&$ 24.69       $&$ 92	 $   		      	      			      		    	      	                                       		     	     \\[1.5pt]\hline	  
                                                      &          &  &          &           &          &         &                               &                                                       &                                                       &                                           &                                                      \\[-12pt]
\rule{0pt}{3ex}\multirow{3}{*}{J092837$+$602521}	&$ 0.1537 $&$ 	8  $&$ 	-3.5    $&$ -14.7      $&$ 14.82       $&$ 38	 $ & \multirow{3}{*}{$0.1540$} & \multirow{3}{*}{$<0.111$} & \multirow{3}{*}{$<14.06$} &\multirow{3}{*}{$\cdots$}&\multirow{3}{*}{$\cdots$}  \\
			  $ $	      			&$ 0.1542 $&$ 	8  $&$ 	30.2    $&$ -12.1      $&$ 19.19       $&$ 48	 $   		      	      			      		    	      	                                       		     	     \\
			  $ $	      			&$ 0.1540 $&$ 	8  $&$ 	67.2    $&$ -12.3      $&$ 35.38       $&$ 89	 $   		      	      			      		    	      	                                       		     	     \\[1.5pt]\hline	  
                                                      &          &  &          &           &          &         &                               &                                                       &                                                       &                                           &                                                      \\[-12pt]
\rule{0pt}{3ex}\multirow{2}{*}{J113327$+$032719}	&$ 0.2367 $&$ 	8  $&$ 	4.5     $&$ -1.7       $&$ 4.79        $&$ 18	 $ & \multirow{2}{*}{$0.2365$} & \multirow{2}{*}{$<0.067$} & \multirow{2}{*}{$<13.79$} &\multirow{2}{*}{$\cdots$}&\multirow{2}{*}{$\cdots$}  \\
			  $ $	      			&$ 0.2364 $&$ 	8  $&$ 	-4.1    $&$ -9.6       $&$ 10.39       $&$ 35	 $   		      	      			      		    	      	                                       		     	     \\[1.5pt]\hline
                                                      &          &  &          &           &          &         &                               &                                                       &                                                       &                                           &                                                      \\[-12pt]
\rule{0pt}{3ex}\multirow{3}{*}{J113910$-$135043}	&$ 0.3598 $&$ 	3  $&$ 	-52.4   $&$ 13.3       $&$ 52.57       $&$ 265   $ & \multirow{3}{*}{$0.3599$} & \multirow{3}{*}{$<0.027$} & \multirow{3}{*}{$<13.36$} &\multirow{3}{*}{$\cdots$}&\multirow{3}{*}{$\cdots$}  \\
							&$ 0.3604 $&$ 	3  $&$ 	-28.0   $&$ 51.7       $&$ 58.41       $&$ 296	 $   		    	      			      		    	      	                                     		     	     \\
							&$ 0.3595 $&$ 	3  $&$ 	-45.9   $&$ 38.5       $&$ 58.95       $&$ 298	 $   		    	      			      		    	      	                                     		     	     \\[-8pt]

	\enddata
	\tablenotetext{a}{Galaxy identification reference: $(1)$ \citet{prochaska11}, $(2)$ \citet{johnson13}, $(3)$ \citet{chen01b}, $(4)$ \citet{chen09}, $(5)$ \citet{muzahid15}, $(6)$ \citet{ellingson91}, $(7)$ \citet{lanzetta95}, $(8)$ \citet{werk12}}
	\tablenotetext{b}{We include known galaxies with $D>350$~kpc for completeness.}
	\tablenotetext{c}{Unless otherwise noted, values were calculated from the {\OVI}~$\lambda 1031$ line.}
	\tablenotetext{d}{Values were calculated from the {\OVI}~$\lambda 1037$ line.}
	\tablenotetext{e}{We quote the $3\sigma$ limits for $W_r$ and $\log N ({\OVI})$}
\end{deluxetable*}

We compare this group environment sample to the isolated galaxy sample presented in \citet{kacprzak15}, who identified 53 isolated galaxies associated with 29 absorbers and 24 non-absorbers. However, the \citet{kacprzak15} sample only extends out to 200~kpc. We supplement this sample with an additional 14 isolated galaxies from the "Multiphase Galaxy Halos" survey to extend the impact parameter range out to 350~kpc. These additional galaxies are presented in Table \ref{tab:iso}. The galaxies in this sample have a median redshift and impact parameter of $\langle z_{\rm gal} \rangle = 0.2608$ and $\langle D \rangle = 259$~kpc, respectively. We find that the galaxies in this extended impact parameter range are all identified as {\OVI} non-absorbers.

\begin{deluxetable*}{llrrrrrrrrr}
	\tablecolumns{11}
	\tablewidth{0pt}
	\setlength{\tabcolsep}{0.06in}
	\tablecaption{Properties of Isolated Galaxies with $200~\text{kpc}~\leq~D~\leq~350~\text{kpc}$\label{tab:iso}}
	\tablehead{
		\colhead{$(1)$}     	&
		\colhead{$(2)$}     	&
		\colhead{$(3)$}     	&
		\colhead{$(4)$}			&
		\colhead{$(5)$}			&
		\colhead{$(6)$}			&
		\colhead{$(7)$}			&
		\colhead{$(8)$}			&
		\colhead{$(9)$}			&
		\colhead{$(10)$}			&
		\colhead{$(11)$}			\\
		\colhead{Quasar J-Name} &
		\colhead{$z_{\rm gal}$} &
		\colhead{RA}     		&
		\colhead{Dec}			&
		\colhead{REF\tablenotemark{a}}			&
		\colhead{$\Delta \alpha$}	&
		\colhead{$\Delta \delta$}	&
		\colhead{$\theta$}			&
		\colhead{$D$}				&
		\colhead{$W_r$}			&
		\colhead{$\log N(\OVI)$}			\\
		\colhead{}     	&
		\colhead{}     	&
		\colhead{(J2000)} &
		\colhead{(J2000)}	&
		\colhead{}		&	
		\colhead{$(\arcsec)$}	&
		\colhead{$(\arcsec)$}	&
		\colhead{$(\arcsec)$}	&
		\colhead{(kpc)}	&
		\colhead{({\AA})}	&
		\colhead{(kpc)}		}
	\startdata
	J012528$-$000555	&$0.4299	$&$01:25:31.548	$&$-00:06:04.22	$&$1	$&$40.6		$&$-8.3		$&$41.46	$&$233.6$ &$<0.061        $&$<13.74$\\
J022815$-$405714	&$0.1248	$&$02:28:10.244	$&$-40:55:48.24	$&$2	$&$-73.8	$&$86.1		$&$102.53	$&$208.6$ &$<0.019        $&$<13.18$\\
J022815$-$405714	&$0.2804	$&$02:28:11.316	$&$-40:56:36.94	$&$2	$&$-57.7	$&$37.4		$&$57.41	$&$248.1$ &$<0.011        $&$<12.95$\\
J040748$-$121136	&$0.2032	$&$04:07:42.680	$&$-12:11:31.87	$&$3	$&$-86.3	$&$4.8		$&$84.45	$&$282.0$ &$<0.005        $&$<12.61$\\  
J040748$-$121136	&$0.2978	$&$04:07:50.585	$&$-12:12:24.23	$&$3	$&$32.3		$&$-47.6	$&$57.10	$&$257.9$ &$<0.005        $&$<12.61$\\
J040748$-$121136	&$0.4100	$&$04:07:45.851	$&$-12:12:13.82	$&$3	$&$-38.7	$&$-37.2	$&$53.03	$&$291.5$ &$<0.006        $&$<12.69$\\
J100402$+$285535	&$0.2143	$&$10:04:03.600	$&$+28:54:33.00	$&$1	$&$14.8		$&$-62.4	$&$63.72	$&$221.6$ &$<0.020        $&$<13.22$\\
J113910$-$135043	&$0.4007	$&$11:39:12.076	$&$-13:51:37.45	$&$1	$&$20.6		$&$-53.8	$&$57.41	$&$310.6$ &$<0.059        $&$<13.73$\\
J130112$+$590206	&$0.2412	$&$13:01:09.806	$&$+59:03:16.35	$&$1	$&$-46.9	$&$69.6		$&$73.67	$&$279.9$ &$<0.017        $&$<13.15$\\
J131956$+$272808	&$0.6717	$&$13:19:59.582	$&$+27:28:06.86	$&$1	$&$50.3		$&$-1.4		$&$44.63	$&$316.2$ &$<0.044        $&$<13.59$\\
J135704$+$191907	&$0.5293	$&$13:57:03.470	$&$+19:18:19.59	$&$1	$&$-14.4	$&$-47.8	$&$49.67	$&$313.8$ &$<0.098        $&$<13.99$\\
J170441$+$604430	&$0.1877	$&$17:04:51.297	$&$+60:44:38.66	$&$1	$&$148.8	$&$8.2		$&$73.19	$&$230.9$ &$<0.031        $&$<13.42$\\
J170441$+$604430	&$0.2260	$&$17:04:31.779	$&$+60:44:42.73	$&$1	$&$-144.0	$&$12.2		$&$71.41	$&$259.9$ &$<0.022        $&$<13.26$\\
J182157$+$642037	&$0.2110	$&$18:22:02.759	$&$+64:21:38.92	$&$4	$&$81.7		$&$62.6		$&$71.85	$&$248.4$ &$<0.005        $&$<12.63$\\[-8pt]

	\enddata
	\tablenotetext{a}{Galaxy identification reference: $(1)$ \citet{chen01b}, $(2)$ \citet{chen09}, $(3)$ \citet{johnson13}, $(4)$ \citet{stocke13}}
\end{deluxetable*}

\subsection{Modeling}
\label{sec:qso_spec}
We performed a  Voigt profile analysis using the software VPFIT\footnote{\url{<https://www.ast.cam.ac.uk/~rfc/vpfit.html>}} \citep{carswell14} to fit the absorption profiles of the {\OVI} absorbers. As the COS instrument on the \textit{HST} has a non-Gaussian line spread function (LSF), leading to significant wings which are prominent at short wavelengths, we use the non-Gaussian LSF from \cite{kriss11}. The LSF for each absorption profile was calculated by interpolating the LSF from \cite{kriss11} to the central wavelength of the profile. This LSF was then convolved with the model profile in the fitting process. The velocity resolution of the FUSE observations is $\sim15-20$~{\kms} (FWHM) and we assumes a Gaussian LSF.

Where possible, both lines of the {\OVIdblt} doublet were used to fit the absorption profile to better constrain the number of components required and determine any blends that might be present. The number of components was based on the minimum number required to produce a reasonably minimized chi-squared. We present the results of the fitting in the second and fourth columns of Figure~\ref{fig:fitplot}. The black line represents the data, while the red line is the model of the {\OVI} absorption profile made using VPFIT. Red ticks indicate the central position of each component used to fit the {\OVI} absorption profile. Three absorbers were found to have blends present. In these cases, we used the unblended transition to fit the profile. The blended transition was then modeled by adding generic components.The total fit in these two instances is shown as a blue line.

From the fitting process, we extracted the column density, Doppler $b$ parameter and redshift of each component of the absorption profiles. We then calculated the equivalent width by integrating the absorption features between the points where the model deviated from the continuum, which we defined as the edges of the absorption profile. These were converted into velocity boundaries using $z_{abs}$, which is the median velocity of the apparent optical depth distribution of the absorption and defines the velocity zero point \citep{cwc-thesis}. The regions contained by these velocity boundaries are shown as the shaded regions in Figure~\ref{fig:fitplot}. We list the equivalent width, column density, $z_{\rm abs}$ and the absorption velocity range in Table~\ref{tab:group}.

We also calculated $3\sigma$ upper limits on the equivalent widths and column densities of the non-absorbing galaxies and group environments. To compute the column density limits, we assumed a single cloud with a Doppler $b$ parameter of $30$~{\kms}, which we found to be typical of {\OVI} absorption in our sample. The equivalent width and column density limits for group environments and isolated galaxy are shown in Tables \ref{tab:group} and \ref{tab:iso} respectively.

\section{Results}
\label{sec:results}
Here we present an analysis of the equivalent width, covering fractions and kinematics of both group and isolated galaxies to determine whether environment plays a role in shaping the properties of the CGM.

\subsection{Equivalent Width and Covering Fraction}
\begin{figure}[ht]
	\includegraphics[width=\linewidth]{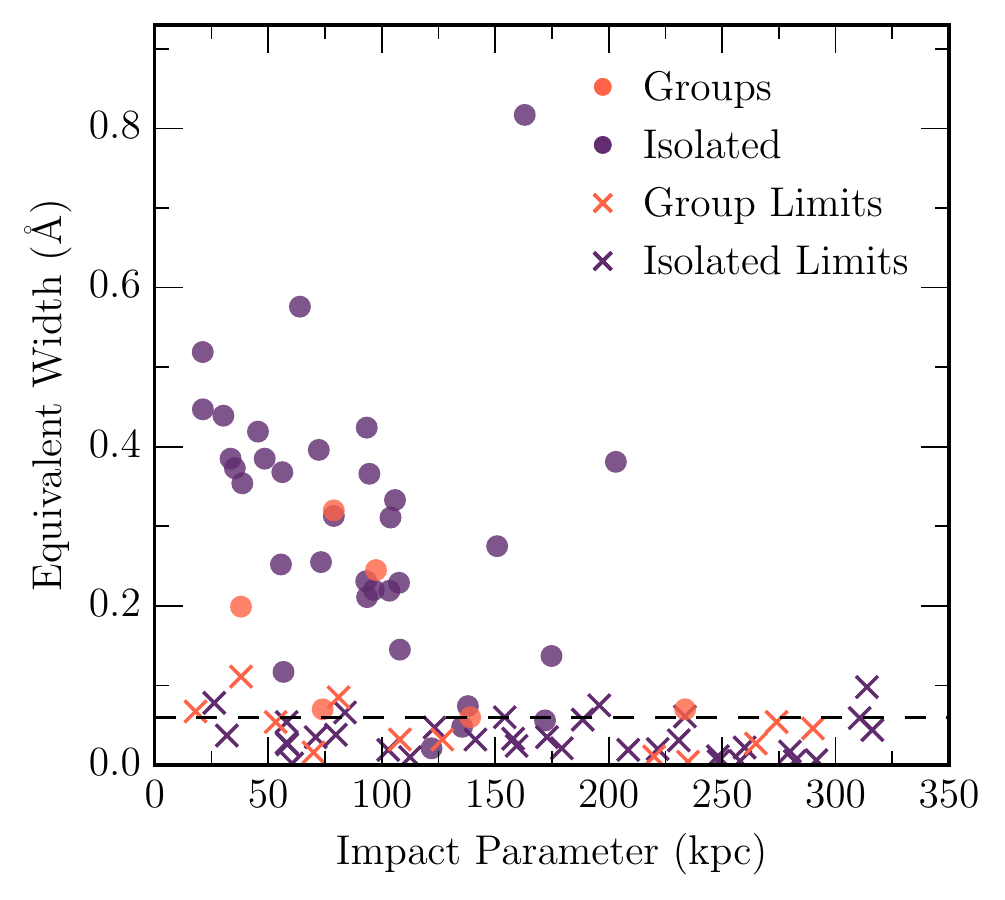}
	\caption[]{Equivalent widths of each {\OVI}~$\lambda 1031$ profile as a function of impact parameter to the nearest member galaxy. The nearest group galaxies are shown in orange and isolated galaxies in purple. Circles indicate where measurements of the equivalent width were possible while crosses show where only a limit could be obtained. The black dashed line indicates the cut-off between absorbers and non-absorbers $(0.06$~\AA$)$.}
	\label{fig:ewvip}
\end{figure}

\citet{kacprzak15} found that the equivalent width decreases as the distance of the absorber from the galaxy increases for the isolated sample. Similarly, \citet{tumlinson11} and \citet{johnson15} found that the column density decreases as the ratio of the impact parameter and the virial radius increases. We investigate this trend in group galaxies by showing the equivalent width as a function of the impact parameter in Figure~\ref{fig:ewvip}. In group environments, the impact parameter is taken to be the projected distance from the quasar to the nearest galaxy. The nearest group members are shown in orange with circles indicating measurements of the equivalent width and crosses representing limits. In our sample, we detect absorbers with equivalent widths greater than $0.06$~{\AA}, which we choose to be the cut-off between absorbers and non-absorbers, represented as a black dashed line. The nearest group galaxies also appear to exhibit decreasing equivalent widths with increasing impact parameter, similar to the trend found in isolated environments. We also investigated the effect of measuring the impact parameter to the geometric centre and the most massive (luminous) galaxy and found that the variations did not change the relationship significantly. Figure~\ref{fig:ewvip} suggests that the equivalent width of group galaxies is lower than what is expected from isolated galaxies. 

\begin{figure}[ht]
	\includegraphics[width=\linewidth]{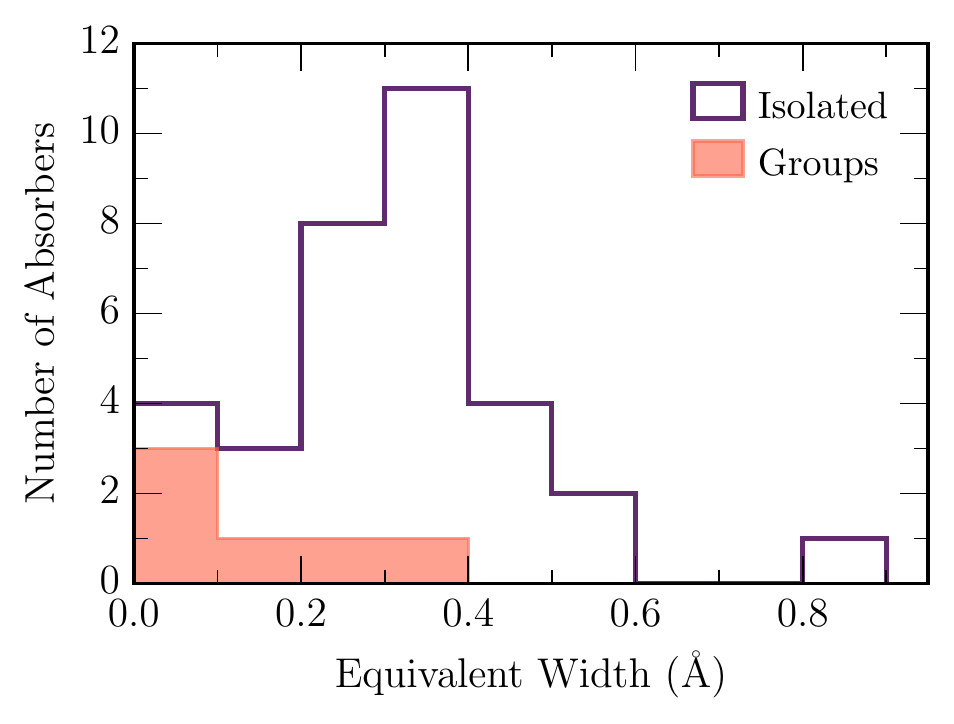}
	\caption[]{The equivalent width distribution of group absorbers (orange histogram) compared to the isolated absorbers (purple histogram). The group absorbers may be offset from the main peak of the isolated absorbers.}
	\label{fig:ewhist}
\end{figure}

We compare the equivalent width distributions of the group and isolated galaxies in Figure~\ref{fig:ewhist}. The peak of the equivalent width distribution of group environments appears to be shifted lower than the peak for isolated galaxies. The average equivalent width for group environments is $\langle W_r \rangle = 0.16 \pm 0.02$~\AA, while the average equivalent width for isolated galaxies is $\langle W_r \rangle = 0.31 \pm 0.03$~\AA. This is a $3\sigma$ difference in the mean equivalent widths. We also tested removing the outlier from the isolated galaxy sample at the top of the plot in Figure~\ref{fig:ewvip} due to its unusually large equivalent width \citep{muzahid15}. The average equivalent width for isolated galaxies then is $\langle W_r \rangle = 0.28 \pm 0.02$~\AA. The difference between the means of the equivalent widths of the group and isolated environments is then $2.4\sigma$. 

The covering fraction, which is the ratio of absorbers to the total number of systems, is an indication of a lack of gas or ideal ionizing conditions in the CGM. We find that the covering fraction for group galaxies is $C_f = 0.33^{+0.06}_{-0.11}$, while the isolated galaxies have $C_f = 0.43^{+0.03}_{-0.06}$, where the $1\sigma$ errors are derived from 10000 bootstrap realizations These are consistent within the errors of the measurements $(0.8\sigma)$, although the smaller covering fraction for the group galaxies suggests that the CGM in these environments appears to have a lack of gas or insufficient ionizing conditions compared to isolated environments. Lower equivalent widths and covering fractions for group galaxies could indicate that these environments are being depleted of {\OVI} and a larger sample would be invaluable to investigate this effect.  

\subsection{Pixel-velocity Two-point Correlation Function}
\label{sec:tpcf_exp}
To compare the kinematics of the {\OVI} absorbers for group and isolated environments, we use the pixel-velocity two-point correlation function (TPCF) method from \citet{magiicat5, magiicat4, nielsenovi}. To construct the isolated galaxy TPCF, \citet{nielsenovi} compiled the pixel velocities within the absorption velocity bounds (i.e., shaded regions in Figure 1) for every isolated absorber in the sample. With this list, they calculated the absolute value of the velocity separations between every pixel pair, binned these velocity separations into 20~{\kms} bins, and normalized the distribution by the total number of pixel pairs in the sample. This produced a probability distribution function, which describes the probability that any two pixels will have a given velocity separation. They used a bootstrap analysis with 100 realizations to obtain $1\sigma$ uncertainties on the TPCF. We use the same method to calculate the group TPCF.

\begin{figure}[ht]
	\includegraphics[width=\linewidth]{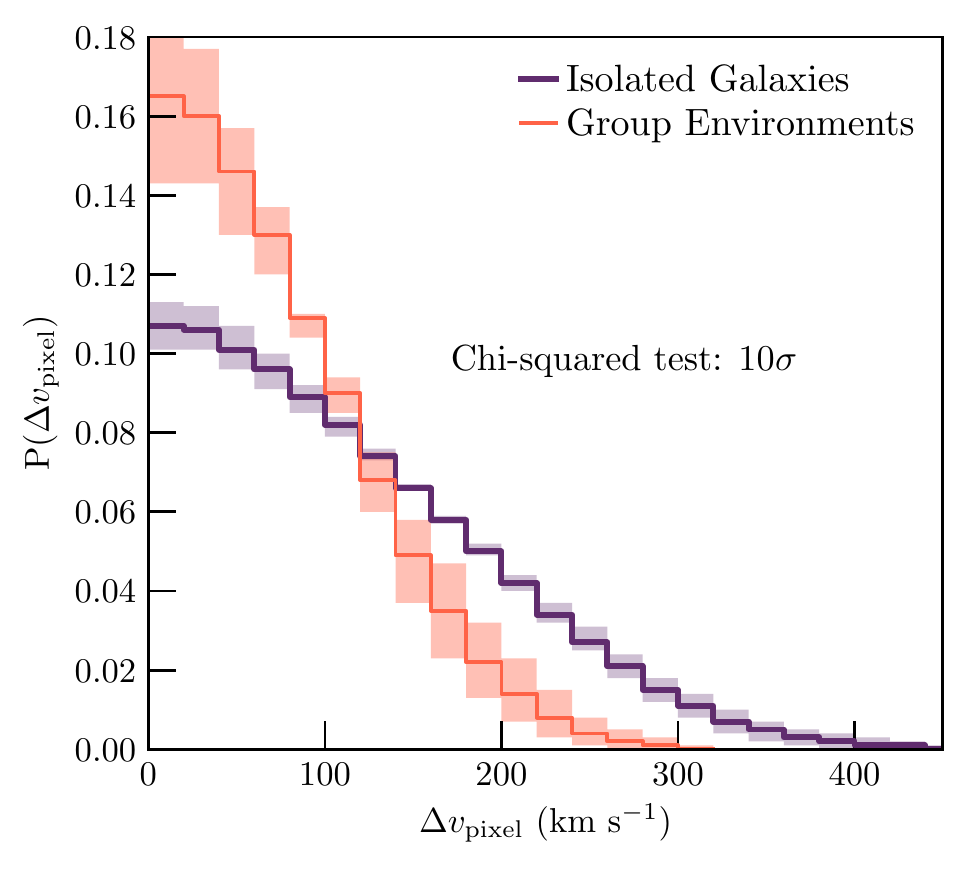}
	\caption[]{The pixel-velocity two-point correlation function (TPCF) comparing the velocity spread of the {\OVI}~$\lambda 1031$ absorption profiles for the group and isolated environments. The group galaxy TPCF is shown in orange while the isolated galaxy TPCF is shown in purple. The result of the chi-squared test is $10\sigma$, indicating that the {\OVI} velocity spread of group absorption profiles is significantly narrower than the isolated TPCF. The shaded region shows the $1\sigma$ errors calculated from bootstrapping.}
	\label{fig:TPCF}
\end{figure}

The group galaxy TPCF (orange) and the \citet{nielsenovi} isolated galaxy TPCF (purple) are shown in Figure 4, where the TPCF is the thick solid line, and the shaded regions represent the one sigma errors calculated from the bootstrap analysis. Any two pixels for a typical group absorption environment are more likely to be closer together in velocity space than two randomly selected pixels from an isolated galaxy. A chi-squared test run on the binned TPCFs, including the uncertainties, found that the null hypothesis that the two samples were drawn from the same population can be ruled out at the $10\sigma$ level. Thus the {\OVI} absorption velocity dispersion is smaller in group environments than in isolated environments.

To characterize the differences in the TPCFs quantitatively, we calculate {\vninety} and {\vfifty} values, which are defined as the velocity separations within which the integrated TPCF curve contains 90\% and 50\% of the data, respectively. Thus, {\vninety} and {\vfifty} are also measures of the velocity spreads of the distributions. We find that {\vninety} is $153_{-18}^{+21}$~{\kms} for group galaxy absorption, compared to $236\pm15$~{\kms} for isolated galaxy absorption ($2.3\sigma$). Similarly, the group absorption {\vfifty} is $ 64_{- 7}^{+ 9}$~{\kms}, while the isolated galaxy absorption is $100\pm6$~{\kms} ($2.4\sigma$). Even though the uncertainties on these values do not overlap, they are still only $2.4\sigma$ different. Thus, combined with the chi-squared test of the TPCF profiles, which is more sensitive to velocity differences in the absorption profiles \citep{magiicat5,magiicat4,nielsenovi}, we find that group environments are more likely to have narrower absorption line profiles than isolated environments. 

\section{Discussion}
\label{sec:discussion}
The UV spectra from \textit{HST}/COS and FUSE have allowed us to compare the {\OVI} CGM in the environment surrounding groups of galaxies to the CGM of isolated galaxies. We have investigated the effects of group environments on the equivalent widths, the covering fractions and the velocity spread of the {\OVI} absorption profiles. 

Differences between the CGM in group and isolated environments have been explored using {\MgII}. \citet{bordoloi11} found that for a given equivalent width, the absorption associated with group galaxies exists out to larger impact parameters than for isolated galaxies. This is supported by \citet{chen10a} who did not find a strong inverse correlation between equivalent width and impact parameter. \citet{bordoloi11} suggests that this is a result of the CGM of group members being superimposed with little interaction between the halos. Nielsen (in prep) found that the velocity spread of absorption profiles for group environments is larger than the velocity spread for isolated galaxies. They also found that a higher fraction of high-velocity components (relative to $z_{\rm abs}$) exist in group halos. This could suggest some interaction occurring between the halos of the group galaxies. We study {\OVI} absorption to investigate the influence of group environments on the warm/hot CGM.

Using the EAGLE simulations, \citet{oppenheimer16} found interesting results with mass that may explain the difference between {\OVI} absorption from the CGM surrounding group and isolated environments. By examining simulated {\OVI} column densities, the authors found that halo mass, and therefore virial temperature, governs the presence and strength of {\OVI}, where sub-L* halos do not have sufficient virial temperatures to create significant amounts of {\OVI}. The most favorable halo masses, and hence virial temperatures, for the creation of {\OVI} are from L* galaxies, which are represented by our isolated sample. As the halo mass increases above that of a typical L$^*$ galaxy to group and cluster environments, the amount of {\OVI} absorption decreases due to the temperatures in the halo preferentially ionizing {\OVI} to higher states, such as {\OVII}, which are difficult to observe. In the case of studying group galaxies, the halo mass is certainly larger that this limit, hence we should start to see this effect. 

The effects of group environments on absorption properties such as covering fractions and equivalent widths were investigated to determine if the trends suggested by \citet{oppenheimer16} were visible in our sample. Taking these results into account, we expected that the equivalent widths and covering fractions for our group sample would be lower than for isolated sample if the  virial temperatures were higher in the group sample. The average equivalent width of the group sample is lower than the isolated galaxies $(3\sigma)$. However, the covering fractions of the absorption associated with group environments, while lower than what were found for isolated galaxies, are consistent with the values from \citet{kacprzak15} $(0.8\sigma)$. This suggests that the mass trends with the presence of {\OVI} could be present but are limited by the small number of group systems in this study. The unchanged covering fraction for group samples may be due to interactions between the group members stripping gas from the galaxies into the CGM. Further investigations should focus on improving the sample size of group environments to determine if groups environments have significantly smaller covering fractions.

The effects of group environments on {\CIV} was investigated by \citet{burchett16} who found that groups with more galaxy members have a deficiency of {\CIV} absorption. Similarly, we find that the larger groups of galaxies tend to have little or no {\OVI} absorption. In our study, the largest group galaxy environment with 8 members, J022815$-$405714 at $z_{abs} = 0.2677$, had an upper limit of $W_r = 0.016$~{\AA} and was classed as a {\OVI} non-absorber. Additionally, the second largest group with 5 members, J040748$-$121136 at $z_{abs}= 0.0919$, was a weak {\OVI} absorber with $W_r =  0.08$~{\AA}. Thus, it is possible that the gas traced by {\OVI} in dense group environments could also be undergoing ram pressure or tidal stripping, leading to the gas being removed from the CGM. Future studies could focus on building a complete sample of groups with a range of member galaxies to investigate this trend.

While the effect of group environments on the CGM was less evident in the covering fractions, we further support the differences in the average equivalent width of the two samples with the kinematics, which were found to be significantly different from isolated environments. The TPCFs showed that, on average, group environment absorption is narrower than isolated galaxy absorption with a difference between the two populations of $10\sigma$. To quantify this, {\vninety} for the group environments $(153^{+21}_{-18}$~{\kms}) is roughly two-thirds that of the isolated sample ($236\pm15$~{\kms}). \citet{stocke14} similarly found that the total absorption profiles of {\OVI} absorbers were narrower for group environments. However, the individual clouds within the profiles were represented by broader components which indicates higher virial temperatures. This narrowing of the absorption profiles, along with potentially smaller average equivalent widths and covering fractions, strongly suggests that the warm CGM does not exist as a superposition of halos as found in {\MgII} by \citet{bordoloi11}. This is further supported by \citet{stocke14} who found that {\OVI} absorbers were not associated with the nearest galaxy in the group environment and favored a mixed CGM scenario. If the group environment CGM properties were due to a superposition of halos, we would expect to see similar or larger velocity widths in group environments compared to the isolated galaxies, which we did not observe. 

Similarly to \citet{stocke14}, we suggest that the regions of warm {\OVI} CGM gas would exist at the interface between the higher virial temperature, diffuse medium of the combined group CGM and the cooler gas associated with individual group galaxies. We suggest that the narrow absorption profiles of {\OVI} in group environments is due to ionisation of lower column density gas in the higher velocity wings of the absorbers. Thus, the {\OVI} gas regions are ionised from the outside of the region's mass, effectively forming an 'onion skin' effect with higher ionisation states on the outside and decreasing inwards. This would create a shell of higher ionised oxygen surrounding the {\OVI}, leading to the observed smaller velocity spread. Similar ideas have been found by \citet{shull98,shull03,tripp01,aracil06,churchill12,kacprzak13,stocke14,stocke17} and \citet{stern16}. Thus, narrower profiles could be the result of a higher virial temperature causing regions of {\OVI} to be ionised from the outside inwards. 

The destruction of {\OVI} and its existence at interface regions means that {\OVI} is an interesting probe of the processes which occur in group environments. More information of the diffuse medium surrounding the interface could be found from higher ionisation states such as {\OVII} \citep{shull03}. However, at this redshift, {\OVII} is in the x-ray regime and not detectable with current telescopes. Since {\OVI} accounts for such a small fraction of the total baryons and oxygen budget \citep{peeples14,oppenheimer16} investigating higher ionisation states such as {\OVII}, {\OVIII}, {\NeVIII} and {\MgX} could further the understanding of the total baryon contents of the CGM. 


\section{Summary and Conclusions}
\label{sec:conclusions}

We compare the CGM for a total of six group galaxy environments to the CGM of 29 isolated galaxies from \citet{kacprzak15} associated with {\OVI} absorption. We also have another 12 groups and 38 isolated galaxies with only an upper limit on {\OVI} absorption. Group environments were classified as having at least two galaxies with a line-of-sight velocity separation of less than $1000$~{\kms}, and the nearest member located within 20 to 350~kpc of the quasar sightline. 

We modeled the absorption profiles for the group sample using VPFIT to obtain the column densities, Doppler $(b)$ parameters and redshifts, from which we calculated the equivalent width and velocity spreads. We directly compared the equivalent widths and covering fractions of the two samples, and used the pixel-velocity TPCF method from \citet{magiicat5,magiicat4} to compare the velocity spreads of the group and isolated galaxy samples. Our findings are:

\begin{enumerate}[nolistsep]

  \item The equivalent width of the absorption in group environments follows the previously established trend of decreasing as the impact parameter increases.
Larger samples of group galaxies could determine if an anti-correlation exists similar to that found for isolated galaxies.
  \item The average equivalent width of absorption associated with group environments is $0.16\pm0.02$~{\AA} compared to the larger value for isolated galaxies, which is $0.31\pm0.03$~{\AA}. The difference between the two populations is $3\sigma$, which is significant. 
  \item The covering fraction is $C_f = 0.33^{+0.06}_{-0.17}$ for group galaxies and $C_f = 0.43^{+0.03}_{-0.06}$ for isolated galaxies. The covering fraction of absorption associated with group environments is smaller than that for isolated galaxies, but they are consistent within the errors of the populations with a difference of only $0.8\sigma$. 
  \item The velocity spread of the absorption profiles of group environments was found to be significantly narrower than the velocity spread of the absorption profiles of isolated galaxies using the TPCF analysis, and a Chi-squared test found this result is significant at the $10\sigma$ level. A comparison of the velocity spread between the group sample ({\vninety}~$=154\pm 19$~{\kms}, {\vfifty}~$=66\pm8$~{\kms}) and the isolated sample ({\vninety}~$=236\pm15$~{\kms}, {\vfifty}~$=100\pm6$~{\kms}) found that the difference in these values between the two samples was at least $2.3\sigma$.

\end{enumerate}

The narrow velocity spreads in group environments indicate that {\OVI} exists at boundaries between the more highly ionised, and diffuse CGM and the cooler gas, which is in agreement with \citet{shull98,shull03,stocke14,stocke17}. We find that is it unlikely that the warm and hot CGM exists as a superposition of the galaxy group member halos, unlike the scenario suggested by \citet{bordoloi11} for the cool CGM. This idea is consistent with \citet{oppenheimer16} who found that {\OVI} is strongly dependent on mass as more massive halos have larger virial temperatures, indicating that the CGM in group environments is warm enough to ionise oxygen to {\OVII} or higher. The lower average equivalent width and slightly smaller covering fraction for group environments also suggests that warm gas is being ionised further in group environments.
 
The {\OVI} absorption towards group environments is markedly different from that detected towards isolated galaxies, although we would benefit from a larger sample size to better understand some of the differences. A study of the higher ionisation states of the CGM in group environments would help us to understand the processes which are occurring in group environments. In particular, to investigate the narrowing of the {\OVI} absorption profiles and the non-superposition of warm CGM halos surrounding group galaxies. Such a study is not currently feasible as these features would be found in the x-ray regime. However, investigating other high ionisation states, located in the UV, would also help to broaden our understanding of the CGM in group environments. 

\acknowledgments

Support for this research was provided by NASA through grants HST GO-13398 from the Space Telescope Science Institute, which is operated by the Association of Universities for Research in Astronomy, Inc., under NASA contract NAS5-26555. S.K.P acknowledges support through the Australian Government Research Training Program Scholarship. G.G.K.~acknowledges the support of the Australian Research Council through the award of a Future Fellowship (FT140100933). SM acknowledges support from European Research Council (ERC), Grant Agreement 278594-GasAroundGalaxies. CWC and JC acknowledge the support from award 1517831 from the National Science Foundation.

\bibliographystyle{apj}
\bibliography{refs}

\end{document}